# Repair kinetics of DSB-foci induced by proton and α-particle microbeams of different energies

Ana Belchior[1,*], João F. Canhoto[1,2], Ulrich Giesen[3], Frank Langner[3], Hans Rabus[4], Reinhard Schulte[5]

[1] Centro de Ciências e Tecnologias Nucleares, Instituto Superior Técnico, Universidade de Lisboa, Estrada Nacional 10 (km 139,7), 2695-066 Bobadela LRS, Portugal
[2] Departamento de Física, Instituto Superior Técnico-IST, Universidade de Lisboa-UL, Av. Rovisco Pais, 1049-001 Lisboa, Portugal
[3] Physikalisch-Technische Bundesanstalt (PTB), 38116 Braunschweig, Germany
[4] Physikalisch-Technische Bundesanstalt (PTB), 10587 Berlin, Germany
[5] Division of Biomedical Engineering Sciences, Loma Linda University, Loma Linda, CA 92350, United States of America

E-mail: anabelchior@tecnico.ulisboa.pt

## Abstract

In this work, the induction and repair of radiation-induced 53BP1 foci were studied in human umbilical vein endothelial cells irradiated at the PTB microbeam with protons and α-particles of different energies. The data were analyzed in terms of the mean number of 53BP1 foci induced by the different ion beams. The number of 53BP1 foci found at different times post-irradiation suggests that the disappearance of foci follows first order kinetics. The mean number of initially produced foci shows the expected increase with LET. The most interesting finding of this work is that the absolute number of persistent foci increases with LET but not their fraction. Furthermore, protons seem to produce more persistent foci as compared to α-particles of even higher LET. This may be seen as experimental evidence that protons may be more effective in producing severe DNA lesions, as was already shown in other work, and that LET may not be the best suited parameter to characterize radiation quality

Keywords: radiation-induced foci, track structure, DNA damage repair

## 1. Introduction

In biological systems exposed to ionizing radiation (IR), lesions induced in the DNA molecule are the starting point for the radiobiological consequences. Such lesions comprise single strand breaks (SSBs), modified bases, abasic sites, DNA-protein crosslinks and double strand breaks (DSBs) (Ward 1988). DSBs are the most complex lesions and, therefore, difficult to remove from the genome. In its simplest form, a DSB comprises two SSBs occurring in close proximity on opposite DNA strands. In normal cells, approximately 1 % of SSBs, which frequently occur during normal metabolism, are converted to DSBs resulting in approximately 50 endogenous DSBs per cell per cell cycle that are usually repaired with high fidelity (Vilenchik and Knudson 2003).

A unique feature of radiation-induced DNA lesions is the occurrence of clustered DNA damage sites that are more difficult to repair than the more abundant isolated DNA lesions such as SSBs and base lesions (Pastwa et al 2003).

When a DNA DSB occurs, hundreds of molecules of a variety of DNA damage response protein species accumulate at the DSB sites in large aggregates that can be made visible in microscopy by fluorescence tagging of some of the involved protein species (Du et al 2003). Examples of such assays are phosphorylated histone H2AX foci (γ-H2AX) and tumor suppressor TP53 binding protein 1 (53BP1 foci) (Fernandez-Capetillo et al 2002).

Several studies showed that the induction of radiation-induced foci (RIF) occurs within minutes of IR exposure and peaks around 30 minutes after irradiation (Rogakou et al 1998, Anderson et al 2001, Mosconi et al 2011, Hable et al 2012). Anderson et al (2001) found 53BP1 foci to appear slightly later than γ-H2AX foci and suggested that 53BP1 might be involved in the repair or checkpoint control associated with persistent foci. Therefore, this study focused on the time evolution of 53BP1 foci after exposure to protons and α-particles.

Generally, the complexity and reparability of DNA damage is mostly attributed in literature to linear energy transfer (LET). The accumulated knowledge, reviewed by Georgakilas *et al* (2013), suggests that the level of complexity increases with LET leading to a compromise of reparability. Jezkova et al. (2018) suggested that the level of DNA damage complexity is dependent on the particle track core diameter, revealing that similar LET and energy may generate different types of DNA damage. Higher-LET radiation has a higher number of interactions, which increases the local dose deposition (Gulliford and Prise 2019) and enhances the biological effectiveness of cell killing when compared to low-LET radiation. For low-LET IR, about 30 % to 40 % of the energy deposits in DNA result in complex DNA lesions (Datta *et al* 2006), as compared to 90 % for high-LET IR (Nikjoo *et al* 1999).

In a previous study reported in (Gonon *et al* 2019), the induction of DNA repair foci at 30 minutes after irradiation of the same cell type with the same radiation qualities as used in this study has been investigated. The main aim of this study was to investigate differences in the effectiveness of DNA repair processes after exposing cells to high- and low-LET particle radiation. The study focused on normal Human Umbilical Vein Endothelial Cells (HUVEC) to assess 53BP1 phosphorylation response in healthy tissues in order to study the repair without the interference of cancer-altered signaling pathways.

The cell experiments were performed in the frame of the BioQuaRT project (Rabus *et al* 2014, Palmans *et al* 2015) at the ion microbeam operated at the Physikalisch-Technische Bundesanstalt (PTB) in Braunschweig, Germany (Greif *et al* 2004). The abundance of 53BP1 foci in the irradiated cells was followed over 24 hours after irradiation and quantitatively analyzed based on a first-order kinetic model from literature (Schultz *et al* 2000, Marková *et al* 2007), assuming two classes of foci with different repairability (Plante *et al* 2019). The analysis is based on a developed approach for modeling the irradiation at the ion microbeam (Gonon *et al* 2019).

The present analysis differs from the one presented in (Gonon *et al* 2019) in that the time dependence of the number of foci has been studied to assess the difference in DNA repair. In addition, the measured data for the different radiation qualities were not treated independently. On the contrary, a simultaneous non-linear regression of all datasets of different radiation qualities and unirradiated samples to obtain the average number of induced and persistent 53BP1 foci per ion track. In doing so, a model function was applied that explicitly accounted for the clustering of ion tracks which lead to indistinguishable foci. In this way, the resulting effect on the time evolution of observed foci is included as an important ingredient that goes beyond models estimating mis-counting due to overlapping of foci (Rabus *et al* 2019, Shqair *et al* 2022).

## 2. Materials and Methods

### 2.1 Cell culture

Cultures of primary human umbilical vein endothelial cells (HUVEC) were obtained from Lonza (Bâle; Switzerland). All cells tested negative for mycoplasma, bacteria, yeast, and fungi. Cells at passage 2 were grown in endothelial cell growth media (EBM® and supplements) (Lonza) containing 4.72 % (vol/vol) fetal bovine serum (Lonza), hydrocortisone, hFGF-B, VEGF, R3-IGF-1, ascorbic acid, HEGF, gentamicin and amphotericin-B (EGM-2BulletKit; Lonza). They were maintained at a temperature of 37 ºC in a humidified incubator in an atmosphere containing 5 % $CO_2$ (vol/vol) in air.

### 2.2 Microbeam irradiations

HUVEC cells were irradiated, following the irradiation procedure described in detail in (Gonon *et al* 2019), at the PTB microbeam facility (Greif *et al* 2004). At about 20 h before irradiations, confluent cell cultures were trypsinized and approximately 4,000 cells were seeded onto stainless-steel dishes with a 25-µm-thick hydrophilic bioFoil base. Dishes were maintained, for 2 h, at 37°C in a humidified atmosphere of 5 % $CO_2$. Then, the dishes were filled with fresh culture medium and remained in the incubator overnight. On the irradiation day, cells were stained with a 150 nM solution of Hoechst 33342 dye (AAT Bioquest Inc., Sunnyvale, CA) for 30 minutes.

The cell dishes were positioned perpendicular to the beam on a computer-controlled XY-stage (Märzhäuser, Wetzlar, Germany) mounted on an inverse microscope (Zeiss Axiovert 100; Oberkochen, Germany) (Greif *et al* 2004). Before irradiation, the dishes were scanned using a 20x magnification lens, an LED-based light source emitting near-UV radiation with wavelengths of (399 ± 9) nm, a sensitive CCD camera and a custom-build analysis software for online determination of the nuclei positions. During both the nuclei scan and irradiation, the dishes were maintained at 37°C. Cell cultures were irradiated under conditions described in Table 1. The different radiation qualities are labelled by the particle type and the initial beam energy.

**Table 1.** Energy and LET values of the different types of radiation used for cellular irradiation.

| Particle type and beam energy | Estimated energy at cell nucleus center (MeV) | Estimated LET at cell nucleus center (keV/µm) |
|---|---|---|
| α-particles | | |
| 20 MeV | 17.8 ± 0.2 | 36 ± 1 |
| 10 MeV | 5.5 ± 0.4 | 85 ± 4 |
| 8 MeV | 1.9 ± 0.6 | 170 ± 40 |
| Protons | | |
| 3 MeV | 1.6 ± 0.2 | 19 ± 2 |

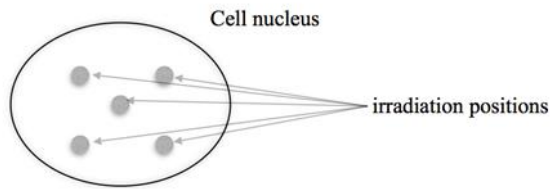

**Figure 1:** Illustration of the targeted irradiation positions for ions in the cell nucleus. The pattern was fixed in space and not adjusted to the nuclei orientation.

The energy at the cell nucleus center and the associated unrestricted linear energy transfer (LET) were determined using the SRIM code (Ziegler *et al* 2010). In these simulations, the passage of the ions through the microbeam exit window, a scintillator foil and the 25 µm-thick bioFoil of the cell-dish was considered, and a thickness of the cell nucleus of 2.4 µm was assumed (Gonon *et al*, 2019).

Each cell nucleus was irradiated with 5 ions in a fixed pattern as schematically shown in Figure 1. The target positions of the 5 tracks were the corners and the center of a 4 µm side square whose orientation was fixed (i.e., the sides of the square were generally not aligned with the axes of the elliptical cell nucleus cross-section).

The actual positions of ion traversal scatter about the target points owing to a beam size of about 4 µm full width at half maximum (Supplementary Table 2). In addition, there is the possibility of zero or two ions being emitted when the detection system at the microbeam registers the passage of an ion. In consequence, the average number of ions traversing a cell nucleus slightly deviates from 5, and two ions may traverse the nucleus in close proximity so that foci produced by their tracks may be indistinguishable. The dimensions of the irradiation pattern were chosen such as to simultaneously minimize both effects (Gonon *et al* 2019).

After irradiation, the dishes were placed back in the incubator. Control cell cultures and sham-irradiated cell cultures were treated as the irradiated ones, but not exposed to IR. The difference between control and sham dishes is that the latter were subject to the nuclei recognition procedure that involved staining with Hoechst 33342 and illumination with light of 399 nm wavelength.

*2.3 Immunostaining and microscopy*

At several times post-irradiation, 0.5 h, 2 h, 4 h, 8 h and 24 h, cells were fixed with 4 % formaldehyde in PBS for 15 minutes, at room temperature, washed with PBS and permeabilized with a 0.5 % Triton X-100 lysis solution, for 3 minutes. After being washed twice with PBS, the cells were again incubated for 1 h with the rabbit polyclonal anti-53BP1 antibody (1/1000, Bethyl A300-272A). Cells were once again washed with BSA 2 % and further incubated for 1 h with the secondary antibody anti-rabbit IgG coupled to Texas Red®-X (1/1000, Invitrogen, T-6391). Finally, the cells were washed three times more with PBS, incubated for 5 minutes with 4',6-Diamidino-2-Phenylindole, Dihydrochloride (14.3 µM, Invitrogen, D1306) and mounted with anti-fade Prolong® Gold (Invitrogen, P36930). Cells were analyzed at 64x magnification in a fluorescence microscope. Image analysis of 53BP1 foci was performed by the freeware Cellprofiler (Carpenter *et al* 2006).

*2.4 Data analysis*

To consider the inter-dish variability and to increase the statistical power of the analysis, three different dishes and a total of 1000 nuclei per dish were analyzed for each radiation quality and time point as well as for sham and control samples.

The data analysis was based on the following assumptions:
- The expected number of foci per nucleus and the respective uncertainty for a radiation condition were estimated, respectively, as the mean and the sample standard deviation of the mean values found in the three replicate experiments.
- The possibility of indistinguishable foci in case of tracks passing the nucleus in proximity was taken into account using a development of the approach of (Gonon *et al* 2019): The mean number of tracks in proximity (leading to indistinguishable foci) were determined by a simulation of the irradiation, separately for each possible number of ions in such a track "cluster". The positions of the points of ion passage through the image plane as well as the lengths of the main axes and orientation of the ellipse representing a cell nucleus were randomly sampled (Supplementary Tables 1 and 2).
- In addition, the possibility that several foci are formed within an ion track and are indistinguishable is taken into account. The number of foci formed in an ion track is assumed to be Poisson distributed.
- It is assumed that radiation-induced foci and foci induced by non-radiation causes occur statistically independently.
- Sham irradiated foci are assumed to be always repairable whereas for radiation-induced foci it is possible that foci are persistent.
- Repair of foci is assumed to be below the saturation point and to follow first order kinetics with a repair rate independent of radiation quality. The first assumption seems justified because even at the highest LET values, the dose to the nucleus from the five passing ion tracks is less than 1 Gy.

The kinetics of 53BP1 foci disappearance was therefore modeled by equations 1 to 4.

$$m_Q(t) = m_s(t) + m_{r,Q}(t) \quad (1)$$

$$m_s(t) = \bar{n}_b + \bar{n}_s e^{-\beta_0 (t-t_M)} \quad (2)$$

$$m_{r,Q}(t) = \sum_{n_i} P_Q(t|n_i)\bar{k}_Q(n_i) \quad (3)$$

$$P_Q(t|n_i) = 1 - e^{n_i \bar{n}_Q \left((1-p_Q)e^{-\beta_1(t-t_M)} + p_Q e^{-\beta_2(t-t_M)}\right)} \quad (4)$$

where $m_Q(t)$ and $m_s(t)$ are the mean number of foci per nucleus observed at time $t$ post irradiation with radiation quality $Q$ and in sham irradiated cells, respectively. $m_{r,Q}(t)$ is the mean number of observed foci produced by ions. $P_Q(t|n_i)$ is the probability of observing a focus at time $t$ post irradiation at the location of a cluster of $n_i$ ions (of radiation quality $Q$) traversing the nucleus in proximity, and $\bar{k}_Q(n_i)$ is the mean number of such clusters of $n_i$ ions.

$\bar{n}_b$ is the mean number of background foci, $\bar{n}_s$ is the mean number of foci per cell nucleus at time $t_M = 0.5$ h due to sham irradiation, and $\beta_0$ is the repair rate of these foci. $\bar{n}_Q$ is the mean number of radiation-induced foci (RIF) formed along an ion track inside the cell nucleus at time $t_M$ post irradiation, $p_Q$ is the fraction of persistent RIF produced by an ion track of radiation quality $Q$, and $\beta_1$ and $\beta_2$ are the repair rates of normal and persistent RIF, respectively.

Alternatively, the probability $P_Q(t|n_i)$ was also modeled by the mathematically equivalent expression

$$P_Q(t|n_i) = 1 - e^{n_i\left((\bar{n}_Q - \bar{p}_Q)e^{-\beta_1(t-t_M)} + \bar{p}_Q e^{-\beta_2(t-t_M)}\right)} \quad (5)$$

where $\bar{p}_Q$ is the mean number of persistent RIF formed along an ion track inside the cell nucleus at time $t_M$ post irradiation.

## 3. Results

### 3.1 53BP1 Foci Background

As described before, prior to irradiation, cells were stained with Hoechst and exposed for about 0.5 s with 399-nm light for nuclei identification. This step may induce additional background foci that are independent of the radiation itself. Therefore, sham-treated samples were prepared in parallel with the irradiated samples. In addition, control samples of cells neither exposed to IR nor 399-nm light were also considered. Figure 2 shows the distribution of the frequency of nuclei counted with $n$ foci for the three control dishes and Figure 3 shows the mean number of foci per nucleus as a function of post-irradiation time for the sham-treated cells. The individual distributions of each experiment on sham-treated cells are presented in Supplementary Figure 1.

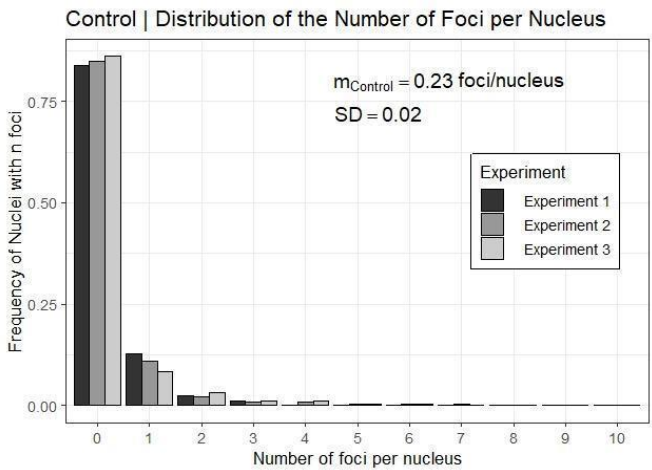

**Figure 2:** Characterization of 53BP1 foci background. Relative frequency distribution of the number of 53BP1 foci per nucleus for three replicate experiments of control dishes. The mean number of foci per nucleus, mControl, represents the mean of the means between the three replicate experiments and its associated standard deviation (SD) computed as the square root of the sample variance between the means.

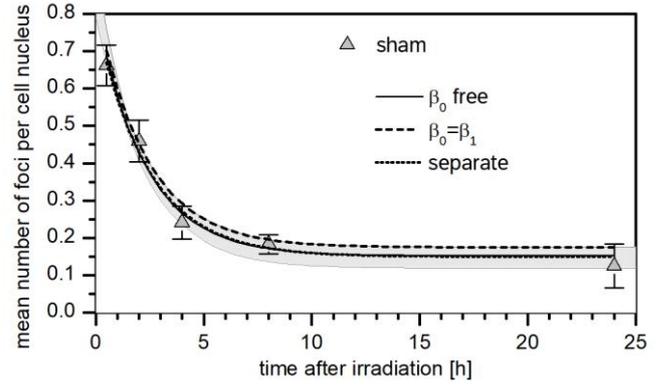

**Figure 3:** Time dependence of the mean number of foci per nucleus as a function of the post-irradiation time for the sham-treated cells. Points represent the mean of means obtained between three replicate experiments for each time point, and error bars represent the SD between the means computed as the square root of the sample variance. The dashed line is the best-fit curve of a regression of only the sham-treated data using equation 2. The solid line is the best-fit curve obtained by simultaneous regression of all data sets; the shaded area indicates the range of results obtained with the different options investigated in the robustness analysis. (See text for details.)

The curves in Figure 3 show the best-fit results of the nonlinear regression. The dashed line represents the best fit obtained when eq. 2 is fitted to the dataset of sham-irradiated cells. The solid line corresponds to the $m_s(t)$ term in eq. 1 which results from a simultaneous regression of all datasets, where the data of the means of all three replicate experiments for the irradiated samples were modeled by equation 1 and the dataset of the means of the three replicates for sham-irradiated cells by equation 2. In the regression, the parameters $\beta_1$ and $\beta_2$ were independent of radiation quality, parameters $\bar{n}_b$ and $\bar{n}_s$ were the same for all datasets, and $\beta_0$ was forced to be equal to $\beta_1$. When the simultaneous regression of all datasets is performed with $\beta_0$ as a free parameter, the resulting fit curve is close to the dashed line in Figure 3. The gray shaded area indicates the range of model values when the different options from the robustness analysis are applied to the entire dataset of irradiated and sham-irradiated cells.

### 3.2. 53BP1 Foci in HUVEC cells targeted with 5 ions

After irradiation, we measured the *in-situ* number of 53BP1 foci, in HUVEC cell cultures fixed at 0.5, 2, 4, 8 and 24 h post-irradiation. The irradiation conditions are described in Table 1 and each nucleus was targeted with 5 particles as illustrated in Figure 1. The distributions of the number of 53BP1 foci per nucleus were evaluated for each time-point, replicate experiment, and radiation quality (Supplementary Figures 2-5). The mean between the means of each experiment and the sample standard deviations for each radiation quality and time point are shown in Figure 4.

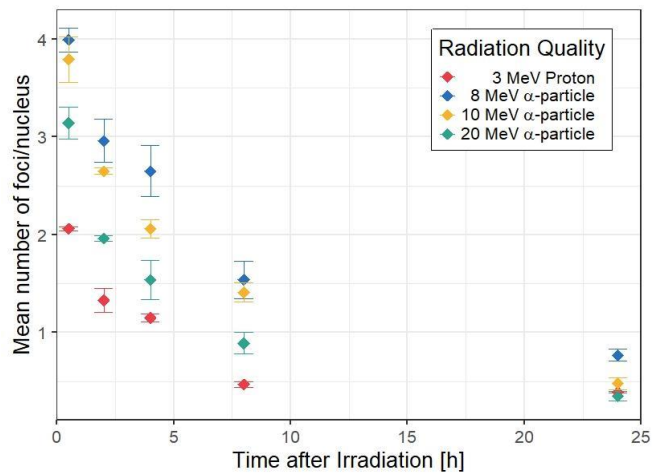

**Figure 4:** Time dependence of the mean number of foci per nucleus for all radiation qualities as a function of post-irradiation time (in hours). Points represent the mean of means obtained between three replicate experiments for each time point, and error bars represent the SD between the means computed as the square root of the sample variance.

### 3.3. Kinetics of the decay of 53BP1 foci

To evaluate the kinetics of the decay of 53BP1 foci, a nonlinear regression analysis was performed for the datasets of all radiation qualities, using equations 1 to 4. Four parameters were determined in this analysis: the mean number of foci per ion track $\bar{n}_Q$ at time $t_M$ after irradiation, the fraction of persistent foci $p_Q$, and the repair rates $\beta_1$ and $\beta_2$. The parameter $t_M$ was set equal to 0.5 h, i.e. the time after irradiation when the maximal number of radiation-induced foci per nucleus is observed (Rogakou *et al* 1998, Anderson *et al* 2001, Mosconi *et al* 2011, Hable *et al* 2012). The model parameters $\bar{k}_Q(n_i)$ were fixed to the values listed in Supplementary Table 3 that were obtained from the Monte Carlo simulations of the irradiations at the ion microbeam. The other model parameters were determined by default with a simultaneous non-linear regression of all datasets in different software environments: Microsoft Excel, R, and GNU data language (GDL). In all cases, the evaluated parameters were constrained to always be positive or equal to zero. The quantity to be minimized was the sum of squares of the ratios of residuals to uncertainties.

In Excel, the solver tool was used with the generalized reduced gradient (GRG) option. The uncertainties of the optimum fit parameters were obtained by calculating (in Excel) the Jacobian and the inverse of the resulting coefficient matrix of the linearized problem. In the analysis in R, the nls()

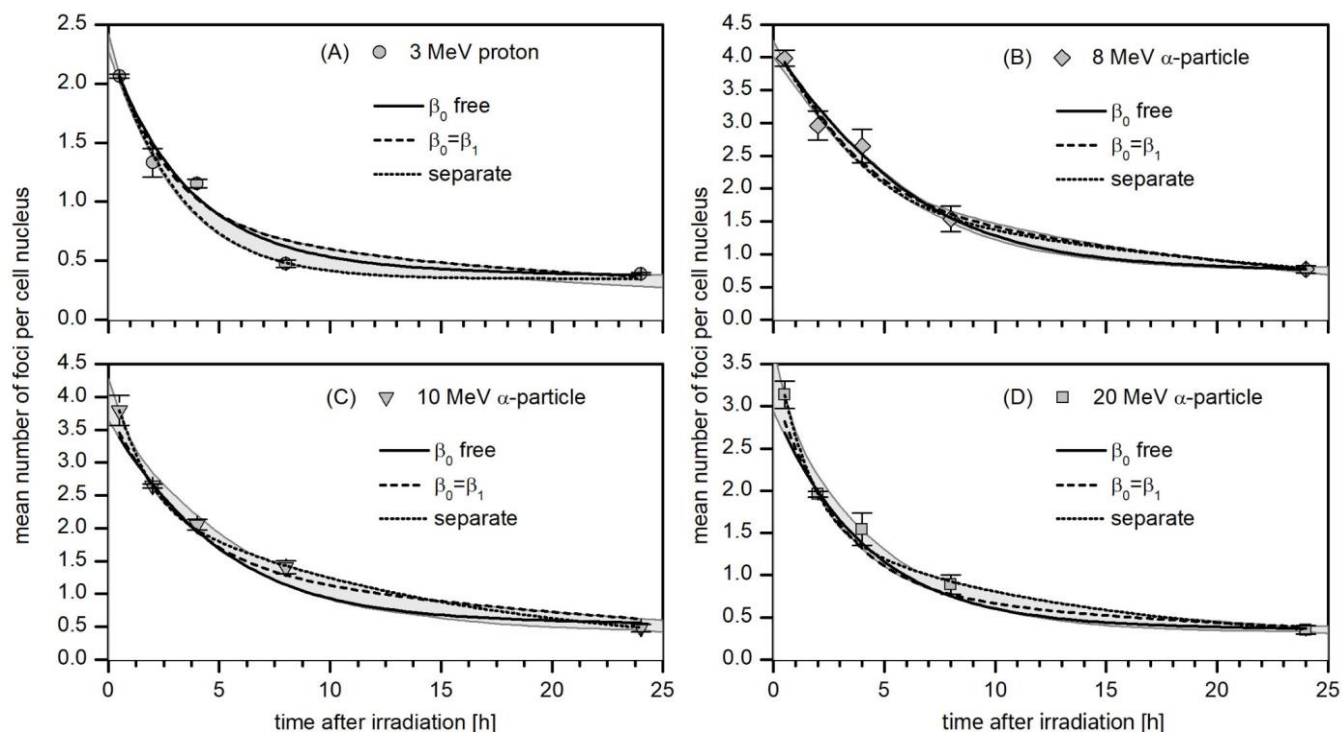

**Figure 5:** Kinetics of the disappearance of 53BP1 foci in HUVEC cells targeted with 3 MeV Protons (A) and 8, 10 and 20 MeV α-particles (B-D). Symbols represent the mean number of foci per nucleus for each time point. The lines are the best-fit curves when the datasets are fitted independently (dotted lines) and when a simultaneous regression of all datasets is performed according to equations 1 to 4 with $\beta_0$ as a free parameter (solid line) or with $\beta_0 = \beta_1$ (dashed lines). The gray shaded areas indicate the range of values obtained from the robustness analysis.

**Table 2.** Results of the model parameters obtained by simultaneous non-linear regression of all datasets: Mean number of radiation-induced foci per track $\bar{n}_Q$, fraction of persistent radiation-induced foci $p_Q$, mean number of persistent radiation-induced foci per track $\bar{p}_Q$, repair rates $\beta_1$ and $\beta_2$, and respective standard errors (SE) obtained from the fit of the non-linear model (equations 1 to 3) to the ensemble of datasets for all radiation qualities and the sham-irradiated cells. The values are from the regression performed using MPfit procedure of GDL using $\beta_0$ in equation 2 as a free parameter. The upper and lower values in the cells in columns 3 through 7 are the fit results obtained by using eq. 4 and eq. 5, respectively, in conjunction with eqs. 1 to 3. The values given in italics in columns 4 and 5 have been calculated from the values in the respective other column. The resulting ratios $\chi^2/f$ of the weighted sum of squared residuals $\chi^2$ (summed over all datasets) to the degrees of freedom f are about 5.4 and 6.2, respectively. In columns 6 and 7 only one value is given, since in the simultaneous fit, these parameters were kept the same for all radiation qualities.

| Radiation Quality | LET (keV/µm) | Mean number of foci per track, $\bar{n}_Q$ | Proportion of persistent foci, $p_Q$ | Mean number of persistent foci per track, $\bar{p}_Q$ | Repair rate, $\beta_1$ ($h^{-1}$) | Repair rate, $\beta_2$ ($h^{-1}$) |
|---|---|---|---|---|---|---|
| protons | | | | | | |
| 3 MeV | 19 ± 2 | 0.37 ± 0.02<br>0.37 ± 0.02 | 0.17 ± 0.12<br>*0.28 ± 0.12* | *0.06 ± 0.04*<br>0.10 ± 0.04 | | |
| α – particles | | | | | | |
| 20 MeV | 36 ± 1 | 0.63 ± 0.04<br>0.63 ± 0.04 | 0.10 ± 0.07<br>*0.16 ± 0.08* | *0.06 ± 0.04*<br>0.11 ± 0.05 | 0.27 ± 0.05<br>0.32 ± 0.07 | 0.01 ± 0.03<br>0.04 ± 0.02 |
| 10 MeV | 85 ± 4 | 1.08 ± 0.06<br>1.09 ± 0.07 | 0.11 ± 0.08<br>*0.21 ± 0.10* | *0.12 ± 0.08*<br>0.23 ± 0.11 | | |
| 8 MeV | 170 ± 40 | 1.66 ± 0.18<br>1.68 ± 0.18 | 0.11 ± 0.08<br>*0.20 ± 0.10* | *0.19 ± 0.14*<br>0.33 ± 0.16 | | |

method was used, and the analysis in GDL was based on the MPfit procedure (Markwardt 2009).

*3.4. Robustness analysis*

To test the robustness of the fit parameters, the non-linear regression was conducted for a variety of start values of the parameters including extreme cases such as all parameters being zero or equal to the maximum possible value (where applicable). Furthermore, the influence of the statistical distribution of the fixed fit parameters $\bar{k}_Q(n_i)$ was also investigated by determining them for 1000 batches of simulated irradiations of 1000 cell nuclei and performing (in GDL) a non-linear regression for each batch. In the simulations, the dimensions and orientations of the cell nuclei and the positions of the ion traversal were randomly sampled. The variation of the parameters $\bar{k}_Q(n_i)$ was found to be in the low percent range for the dominant values of isolated tracks and track clusters with two or three tracks (Supplementary Table 4). The resulting variation of the model parameters determined by fitting different batches was found to be roughly an order of magnitude smaller than the uncertainties of the parameters deduced from the non-linear regression (Supplementary Table 5).

In addition, fits were also performed in which constraints on the parameters were imposed or relaxed such as: $\beta_1$ was allowed to vary between radiation qualities, $\beta_0$ was forced to be identical to $\beta_1$, $\beta_2$ was fixed to 0. Fits to the difference between irradiated cells and sham-irradiated cells (using equation 3 as the model) were also performed.

The four panels (A-D) in Figure 5 show the data of the mean number of foci per nucleus for the four investigated radiation qualities (points), the respective standard deviation and the best-fit curves obtained from the simultaneous regression analysis of all datasets according to eqs. 1 to 4 (solid lines). In addition, the best fit curves from the independent regression of each dataset according to eq. 3 (dotted lines) and from a simultaneous regression with the boundary condition $\beta_0 = \beta_1$ (dashed lines) are also shown. The gray shaded areas indicate the range of fit curve values from the different options in the robustness analysis.

Table 2 summarizes the results obtained by simultaneous regression of all datasets for the mean number of radiation-induced foci per track, $\bar{n}_Q$, the fraction of persistent radiation-induced foci, $p_Q$, and the mean number of persistent radiation-induced foci, $\bar{p}_Q$, for each radiation quality. The last two columns in Table 2 show the repair rates $\beta_1$ and $\beta_2$ that are the same for all radiation qualities. For each of the parameters, two values are listed in Table 2: the upper one is the result from a regression where equation 4 was used for the probability of foci formation in a track cluster, and the lower one is for regression using equation 5. It should be noted that from the model, the parameters $\beta_1$ and $\beta_2$ are identical for all radiation qualities. Therefore, only two values are listed for each of them (depending on whether equation 4 or equation 5 has been used for regression.) Different effectiveness of radiation qualities, therefore, is reflected in the number or proportion of persistent foci only.

**Table 3.** Results of the model parameters obtained by simultaneous non-linear regression of all datasets: Number of radiation-induced foci per track $\bar{n}_Q$, fraction of persistent radiation-induced foci $p_Q$, mean number of persistent radiation-induced foci per track $\bar{p}_Q$, repair rates $\beta_1$ and $\beta_2$, and respective standard errors (SE) obtained from the fit of the non-linear model (equations 1 to 3) to the ensemble of datasets for all radiation qualities and the sham-irradiated cells. The values are from the regression performed using the GDL MPfit procedure imposing $\beta_0 = \beta_1$. The upper and lower values in the cells in columns 3 through 7 are the fit results obtained by using eq. 4 and eq. 5, respectively, in conjunction with eqs. 1 to 3. The values given in italics in columns 4 and 5 have been calculated from the values in the respective other column. The resulting ratio $\chi^2/f$ of the weighted sum of squared residuals $\chi^2$ (summed over all datasets) to the degrees of freedom $f$ is about 4.8 in both cases. In columns 6 and 7, only one value is given since, in the simultaneous fit, these parameters were kept the same for all radiation qualities.

| Radiation Quality | LET (keV/μm) | Mean number of foci per track, $\bar{n}_Q$ | Proportion of persistent foci, $p_Q$ | Mean number of persistent foci per track, $\bar{p}_Q$ | Repair rate, $\beta_1$ ($h^{-1}$) | Repair rate, $\beta_2$ ($h^{-1}$) |
|---|---|---|---|---|---|---|
| protons | | | | | | |
| 3 MeV | 19 ± 2 | 0.37 ± 0.02<br>0.37 ± 0.02 | 0.42 ± 0.06<br>*0.38 ± 0.05* | *0.15 ± 0.02*<br>0.14 ± 0.02 | | |
| α – particles | | | | | 0.43 ± 0.01<br>0.41 ± 0.01 | 0.06 ± 0.01<br>0.05 ± 0.01 |
| 20 MeV | 36 ± 1 | 0.69 ± 0.04<br>0.69 ± 0.04 | 0.25 ± 0.06<br>*0.22 ± 0.05* | *0.17 ± 0.04*<br>0.15 ± 0.03 | | |
| 10 MeV | 85 ± 4 | 1.13 ± 0.06<br>1.13 ± 0.06 | 0.33 ± 0.06<br>*0.30 ± 0.05* | *0.38 ± 0.08*<br>0.34 ± 0.06 | | |
| 8 MeV | 170 ± 40 | 1.68 ± 0.18<br>1.68 ± 0.18 | 0.31 ± 0.08<br>*0.27 ± 0.07* | *0.52 ± 0.15*<br>0.47 ± 0.10 | | |

Table 3 shows the best-fit parameters when light-induced and radiation-induced non-persistent foci are assumed to be repaired with the same kinetics (i.e., $\beta_0 = \beta_1$). The results for the fit parameters from independent fits of the different datasets are listed in Supplementary Table 6.

Observations in Table 2 are that the mean number of radiation-induced foci per track increases with LET. In Table 2 also the mean number of persistent radiation-induced foci per track is seen to increase with increasing LET. The same behavior was found for all fit options considered in the robustness analysis. On the contrary, the proportion of persistent foci is essentially constant for the three α–particle beams, whereas this value is about 50 % higher for protons. When the regression is done with eq. 5, higher values of $\bar{p}_Q$ (and consequently of $p_Q$) are found, while the relative dependence on radiation quality remains the same. The larger $\bar{p}_Q$ values are accompanied by an increased $\beta_2$ that is effectively zero when the regression is done using eq. 4. $\beta_1$ also slightly increases but remains unchanged within the error bar. The $\chi^2/f$ is also increased when eq. 5 is used, which indicates a slightly reduced goodness of fit. However, both values are so large that the model is not explaining the whole variation in the data in both cases.

The full model implicitly assumes a different repair rate for light-induced and fast-repairing radiation-induced foci. If this assumption is abandoned, the values listed in Table 3 are obtained. For these fits, the difference between regression using eq. 4 or eq. 5 appears negligible. This also applies to $\beta_2$ that has values significantly different from 0 in both cases and larger than the values seen in Table 2. Both options lead to a fit with the same $\chi^2/f$ of 4.8, which is slightly reduced compared to the values found with $\beta_0$ as a free parameter. This suggests an improved fit for the ensemble of all datasets, even though the fit to the sham-irradiated data deteriorates, as can be seen in Figure 3.

It is worth noting that when the regression of the different datasets is performed independently, the model curves (dotted lines in Figures 4 and 5) tend to better describe the data. The proportion of persistent foci for protons is then obtained as only a third of those found for the α–particle beams where the absolute number of persistent foci shows no variation within the large error bars (Supplementary Table 6). The repair rates found for this fit option have large uncertainties and don't show a clear trend. This may be related to the fact that the number of data points is as small as 5 which leaves only one degree of freedom for the regression.

Imposing the boundary condition $\beta_2 = 0$ increases the degrees of freedom to 2. Supplementary Table 7 shows the results of the respective separate regression on the data for the different radiation qualities. Generally, the fit parameters appear slightly more consistent than those in Supplementary Table 6. However, the value of $p_Q$ for the 20 MeV α–particle beam is by about a factor of 4 to 6 enhanced with respect to the other two α–particle beams, and for α–particles of 10 MeV beam energy an about 50 % lower fraction of persistent foci is found than protons. This suggests that also with two degrees of freedom the fits are still compromised. It must also be noted that the large spread of results from the different fit options is mainly due to the separate regressions on the individual datasets.

## 4. Discussion

From the results shown in Figures 4 and 5, Tables 2 and 3, and Supplementary Tables 6 and 7, it appears evident that a reliable determination of the repair kinetics is only possible when performing a simultaneous regression of all datasets. This can be related to the fact that only 3000 cells per radiation quality and time point have been analyzed. In the study by Gonon *et al* (2019), where only the appearance of foci was investigated at 30 minutes post irradiation for the same radiation qualities and the same cell line as used here, more replicates were performed and almost an order of magnitude more cells were scored for this initial time point.

In this work, it was necessary to develop the approach used by Gonon *et al* (2019), since the repair kinetics of foci formed in a cluster of tracks deviates from that of isolated ion tracks. For the mean number of foci at 30 minutes after irradiation, it was sufficient (in Gonon *et al* (2019)) to replace the sum in eq. 3 by the product of the total number of track clusters (including "clusters" of just one track) and an average probability for observation of a focus at the location of a track. This fact allows converting the probabilities for foci formation in a track into the mean number $\bar{n}_Q$ of foci formed in a track.

The resulting values of $\bar{n}_Q$ for 3 MeV proton and 20 MeV, 10 MeV and 8 MeV α–particle beams from the work of Gonon *et al* (2019) are 0.27, 0.51, 1.27 and 1.05, respectively, and the value for protons is comparable to that found by Ugenskiene *et al* (2009). Here it must be noted that in the experiments with 8 MeV α–particle beams, Gonon *et al* used the number of foci at 10 minutes after radiation as the biological endpoint. Since the maximum of 53BP1 foci occurs at about 30 minutes after irradiation (Rogakou *et al* 1998, Anderson *et al* 2001, Mosconi *et al* 2011, Hable *et al* 2012), it is expected that the value is significantly smaller than for 30 minutes post irradiation.

The other values are about 20 % smaller than the values reported in Tables 2 and 3 of this work. At the same time, the mean number of foci per cell in control samples as well as in sham irradiated cells have comparable values to what was found in this work ($m_{Control} = 0.24 \pm 0.48$, $m_{Sham} = 0.50 \pm 0.86$).

The present analysis differs from the one presented in Gonon *et al* (2019) in the following respects: On the one hand, a slightly different (but supposedly more realistic) assessment of probabilities for the number of tracks in a cell nucleus has been performed. Furthermore, the small effect of doubly targeted nuclei is ignored. From the analysis in (Canhoto 2018), this is only a small effect on the order of 2 %. In contrast to Gonon *et al* (2019), no discrimination of cell cycle has been done.

A potential explanation for the smaller number of foci per track could be that the average nuclear cross-sectional areas found by Gonon *et al* are about 15 % larger than those found in this work (Canhoto 2018, Belchior *et al* 2019). Further analysis showed that the distributions of the short and long half-axes of the equivalent ellipses in the present work had smaller mean values and, at the same time, larger standard deviations in the principal components of the bivariate distribution of long and short half-axes. The standard deviation of the smaller principal component is 50 % higher than that reported in (Gonon *et al* 2019). For the larger principal component, the difference is even by a factor of 2.

Since the DNA content in the nuclei should be constant for cells in G0/G1 phase, a smaller nucleus cross-sectional area should be related to a longer effective track length in DNA medium and, hence, a larger number of foci along the track segment through the nucleus. In fact, a signature of radiation quality should not be the absolute number of foci produced in an ion track segment, but rather the number of foci produced per unit path length.

A comprehensive investigation of these issues is beyond the scope of this paper and requires further analysis of not just foci counts but also other parameters such as the geometry of foci and cell nuclei, variation of track length, etc. The relevant quantity is the density of foci per track length not their number per nucleus. This requires a more sophisticated data science approach that will be elaborated elsewhere (Canhoto *et al* in preparation).

Returning to the present investigation, the simultaneous analysis of all datasets can be expected to be more robust in any case, since a variation of repair rates and of the fractions of foci following different repair kinetics is giving the model too much freedom and less specificity. Furthermore, subtracting the results from sham irradiated cells from those of the different radiation qualities introduces a correlation between the data to be fitted that is not accounted for in the independent regressions of the different datasets.

One important observation from the present investigation is that the mean number of foci per cell in sham irradiated cells at 24 hours after irradiation is consistently lower than the mean number of foci found in control samples (cf. Figure 2 and Supplementary Figure 1). This suggests an increased production of background foci simply from bringing the cell to the microbeam environment.

However, the most interesting finding of this work is that not the fraction of persistent foci increases with LET but their absolute number. Furthermore, protons seem to produce more persistent foci as compared to α–particles of even higher LET. This may be seen as experimental evidence that protons may be more effective in producing severe DNA lesions as was already shown in other work and that LET may not be the best suited parameter to characterize radiation quality (Rucinski *et al* 2021).

Generally, the complexity and reparability of DNA damage is mostly attributed in literature to LET. The accumulated knowledge, reviewed by Georgakilas *et al* (2013), suggest that the level of complexity increases with LET leading to a compromise of reparability. Jezkova *et al* (2018) suggested that the level of DNA damage complexity is dependent on the particle track core diameter, revealing that similar LET and energy may generate different types of DNA damage. Higher-LET radiation has a higher number of interactions, which increases the local dose deposition (Guilliford and Prise 2019) and enhances the biological effectiveness of cell killing when compared to low-LET radiation.

This enhancement in biological effectiveness is represented by the relative biological effectiveness (RBE). The origin of different RBE lies in the microscopic pattern of energy deposition events, the particle track-structure, where IR interactions in nanometric sites are related to the spatial distribution of DNA damage (Goodhead 1989). Schulte *et al* (2008) observed that for high-LET IR intermediate (3-10 ions) and large (>10 ions) ionization clusters in nanometric volumes corresponding to a short DNA segment were more frequent than small clusters (1 or 2 ions), while the opposite was true for low-LET IR.

Several phenomenological models are aiming to find a better relationship between RBE and LET to better model the biological effectiveness of protons, such as the models of Carabe-Fernandez *et al* (2007) and Wedenberg *et al* (2013). However, as Underwood (2019) states, while LET may be a good representation for the DNA damage complexity, it fails to provide a good overview of energy deposition on the nanoscale. In agreement, a review by Rucinski et al (2021) described that LET does not provide information on the stochastic distribution of energy transfers.

Beyond this fundamental question of which physical properties of radiation track structure are most relevant for the severity of DNA damage and the kinetic of its repair, the question may also be raised about what an exponential time dependence of the number of radiation-induced foci actually means. Such models have been used by several authors in literature (Schultz *et al* 2000, Marková *et al* 2007, Plante *et al* 2019). However, DNA damage response foci are not decaying spontaneously and instantaneously like unstable nuclei or excited atoms. These foci are also not (as a whole) a reaction partner in a single biochemical reaction where the concentration of one reaction partner dominates the number of reactions occurring (as is the case for chemical reactions of first order kinetics). In a simplistic view, any time dependence of the number of observed foci would suggest that the repair time (time until disappearance) has this distribution. Then the question arises to which factors is the repair time related? Could this be, e.g., to foci size? These questions warrant further investigation and will be addressed in a follow-up paper.

A first model investigation has been conducted that assumes that foci are repaired within a certain time interval (of a length that has a distribution between two limiting values) either sequentially one by one or in parallel at a reduced repair rate that depends on the number of foci present. Preliminary results favor the first option and indicate that with such a model a time dependence is obtained that may be indistinguishable from an exponential when only a small number of time points is considered. This encourages the use of more time points in future studies of the disappearance of radiation-induced foci.

## 5. Conclusions

In this work, 53BP1 foci induced by proton and α–particle beams of different LET in human umbilical vein endothelial cells were studied at different time points after irradiation. The scored foci numbers were quantitatively analyzed by regression with a model assuming two classes of DNA damage with different average repair time. While the absolute number of foci and of persistent foci showed the expected increase with LET, the fraction of persistent foci was independent of LET for the three α–particle beams studied. For the lower LET proton beam, this fraction was even higher than for the α–particles. This corroborates previous evidence that LET is not the decisive parameter for biological effectiveness and that protons may be more effective in producing radiation damage than heavier ions of the same LET. Since protons and heavier ions are increasingly used in radiation therapy, with protons being the less expensive modality, these findings may have practical implications for the future development of radiotherapy.

A limitation of the present study is the small number of time points, which results in comparatively large uncertainties of the determined model parameter values. Furthermore, only one beam of protons has been used so no conclusion can be drawn regarding the LET dependence of the fraction of persistent foci for protons. To better understand and corroborate the surprising findings of this work, it is planned to perform further experiments using proton beams of different LET. Further aspects to be considered in future studies include measuring additional endpoints and studying the correlation of the number of total and persistent DNA repair foci with the yield of such endpoints or with other indicators of radiation effects, such as the concentration of radiation-induced reactive species. Augmenting the data analysis model by data science approaches that consider not only foci count but also features of the foci is also expected to give deeper insight into the radiation-quality dependence of DNA damage.

**Acknowledgements**

This work was partially funded by the research grant SIB06-REG3 of the European Metrology Research Program (EMRP) in the framework of the BioQuaRT project. The EMRP was jointly funded by the European Union and the EMRP-participating countries. João F. Canhoto acknowledges financial support from Fundação para a Ciência e a Tecnologia (FCT) through the research grant PRT/BD/151544/2021. Géraldine Gonon is acknowledged for advice on the cell experiments. A. Heiske, T. Klages and J. Rahm for their support with the irradiation of cells and O. Döhr, H. Eggestein, T. Heldt and M. Hoffmann for the operation of the PTB ion accelerators.

**Author Contributions**

Conceptualization, A.B.; formal analysis, H.R.; funding acquisition, H.R.; investigation, A.B.; methodology, A.B. and H.R.; resources, A.B., F.L. and U.G.; validation, J.C.; visualization, H.R and J.C.; writing – original draft preparation – A.B., H.R. and J.C.; writing – review and editing, H.R., J.C., U.G and R.S.

## Supplementary Information

**Supplementary Table 1.** Summary of the approach used in the simulation of ion track clustering that results in indistinguishable foci: The scatter of the actual ion position around the target position was sampled from a two-dimensional Gaussian distribution. The parameters of this distribution have been established experimentally and are given in Supplementary Table 2. Similar to Gonon *et al* (2019), two ion tracks were considered as part of a cluster when their distance was below a set value (2 µm). Different from the approach of Gonon *et al* (2019), the occurrence of zero or two ions emitted when the fluorescence detection system at the microbeam registers the passage of an ion has been included in this simulation and the ion pairs were further analysed to identify clusters of more than two ion tracks. The cell nuclei were assumed to be of elliptical shape. In contrast to (Gonon *et al* 2019), it was taken into account that the half-axes found in the measured nuclei are correlated. Therefore, a principal components analysis (PCA) of the bi-variate distribution of long and short half axes was performed to obtain independently distributed variables that were found to be well approximated by Gaussian distributions. Therefore, the principal components c1 and c2 were randomly sampled from these Gaussian distributions and the long and short half axes were obtained as $\mathbf{a} = \bar{a} + v_1 c_1 - v_2 c_2$ and $\mathbf{b} = \bar{b} + v_2 c_1 + v_1 c_2$, where $(v_1, v_2)$ is the first eigenvector of the PCA and $\bar{a}$ and $\bar{b}$ are the mean values of the distributions of long and short half axes observed with the scored cell nuclei.

| Input quantity | Sampling distribution |
|---|---|
| cell nucleus properties | |
|     equivalent ellipse half axes *a* and *b* | calculated from sampled principal components |
|     principal components of bi-variate distribution of *a* and *b* | Gaussian |
|     equivalent ellipse orientation (azimuth angle of long axis) | uniform between 0 and $\pi$ |
| ion microbeam properties | |
|     deviation from target position | two-dimensional Gaussian |
|     ion passage detection | fixed values |

**Supplementary Table 2.** Parameters used in the simulation of ion track clustering that results in indistinguishable foci. Similar to Gonon *et al* (2019), two ion tracks were considered as part of a cluster when their distance was below a set value.

| Radiation quality | protons 3 MeV | α-particles 20 MeV | α-particles 10 MeV | α-particles 8 MeV |
|---|---|---|---|---|
| beam | | | | |
|     horizontal FWHM / µm | 4.8 | 4.2 | 4.5 | 4.5 |
|     vertical FWHM / µm | 4.8 | 3.9 | 3.5 | 3.5 |
|     noise event frequency [a] | 1 % | 1 % | 1 % | 0.1 % |
|     double event frequency [b] | 1 % | 1 % | 1 % | 1 % |
| cell nuclei | | | | |
|     mean long half axis / µm | 7.60 | 8.31 | 8.28 | 7.73 |
|     mean short half axis / µm | 5.27 | 5.54 | 5.54 | 5.53 |
|     sigma principal component 1 / µm | 1.65 | 1.20 | 1.27 | 1.23 |
|     sigma principal component 2 / µm | 0.61 | 0.53 | 0.55 | 0.55 |
|     first eigenvector | (0.880,0.475) | (0.923,0.385) | (0.909,0.417) | (0.897,0.442) |

[a] Noise event: the fluorescence detection system at the microbeam records the passage of an ion while no ion is emitted

[b] Double event: An ion passes the exit window of the microbeam producing a fluorescence signal below threshold, such that when the fluorescence detection system records the passage of an ion, in reality two ions have been emitted

**Supplementary Table 3.** Simulation results for the mean number of clusters of tracks that result in indistinguishable foci. The simulations were performed for the three α-particle beams with $10^6$ random samples of the positions of the ion trajectories and of nuclei dimensions and orientations. The values for protons are the averages of the three values for the α-particles. Similar to Gonon *et al* (2019), two foci were assumed to be indistinguishable when their distance was below 2 µm.

| Radiation quality | protons 3 MeV | α-particles 20 MeV | α-particles 10 MeV | α-particles 8 MeV |
|---|---|---|---|---|
| isolated tracks | 3.0371 | 3.0435 | 3.0312 | 3.0366 |
| cluster of 2 tracks | 0.5916 | 0.5890 | 0.5910 | 0.5949 |
| cluster of 3 tracks | 0.1500 | 0.1491 | 0.1496 | 0.1515 |
| cluster of 4 tracks | 0.0305 | 0.0305 | 0.0302 | 0.0307 |
| cluster of 5 tracks | 0.0044 | 0.0046 | 0.0041 | 0.0045 |
| cluster of 6 tracks | 0.0001 | 0.0001 | 0.0002 | 0.0001 |

**Supplementary Table 4.** Simulation results for the mean number of clusters of tracks that result in indistinguishable foci. The simulations were performed for the three α-particle beams with $10^6$ random samples of the positions of the ion trajectories and of nuclei dimensions and orientations. Similar to Gonon *et al* (2019), two foci were assumed to be indistinguishable when their distance was below 2 µm.

| Radiation quality | α-particles 20 MeV | α-particles 10 MeV | α-particles 8 MeV |
|---|---|---|---|
| isolated tracks | 3.042 ± 0.029 | 3.039 ± 0.044 | 3.033 ± 0.044 |
| cluster of 2 tracks | 0.586 ± 0.021 | 0.588 ± 0.020 | 0.594 ± 0.021 |
| cluster of 3 tracks | 0.150 ± 0.012 | 0.149 ± 0.011 | 0.152 ± 0.012 |
| cluster of 4 tracks | 0.0312 ± 0.0057 | 0.0306 ± 0.054 | 0.0313 ± 0.0054 |
| cluster of 5 tracks | 0.0044 ± 0.0021 | 0.00042 ± 0.00021 | 0.0044 ± 0.0002 |
| cluster of 6 tracks | 0.00018 ± 0.00043 | 0.00015 ± 0.00039 | 0.00018 ± 0.00043 |

**Supplementary Table 5.** Results for the mean and standard deviation of the number of clusters of tracks that results in indistinguishable foci for α-particles of 20 MeV energy. Similar to Gonon *et al* (2019), two foci were assumed to be indistinguishable when their distance was below 2 µm. The values in the second column are the mean and the sample standard deviation among the means of the free model parameters obtained by non-linear regression using values for the parameters $\bar{k}_Q$ determined from 1000 batches containing each 1000 samples of the positions of the ion trajectories and of nuclei dimensions and orientations. The last column shows the fit results and estimated uncertainties of the for the free model parameter from the last batch. The uncertainties given in this column are representative for the uncertainties due to the non-linear regression for a particular set of parameters $\bar{k}_Q$. The uncertainties from the scatter of the parameters $\bar{k}_Q$ are negligible compared to the uncertainties from the fit procedure.

| parameter | all batches | last batch |
|---|---|---|
| $\bar{n}_{Q,1}$ | 0.403 ± 0.003 | 0.404 ± 0.044 |
| $\bar{n}_{Q,2}$ | 0.647 ± 0.005 | 0.645 ± 0.089 |
| $\bar{n}_{Q,3}$ | 1.124 ± 0.012 | 1.11 ± 0.14 |
| $\bar{n}_{Q,4}$ | 1.769 ± 0.027 | 1.80 ± 0.44 |
| $p_{Q,1}$ | 0.1336 ± 0.0009 | 0.13 ± 0.25 |
| $p_{Q,2}$ | 0.0724 ± 0.0006 | 0.07 ± 0.14 |
| $p_{Q,3}$ | 0.0789 ± 0.0013 | 0.08 ± 0.15 |
| $p_{Q,4}$ | 0.0811 ± 0.0014 | 0.08 ± 0.16 |
| $n_b$ | 0.1459 ± 0.0001 | 0.15 ± 0.09 |
| $n_s$ | 0.5464 ± 0.0004 | 0.55 ± 0.15 |
| $\beta_0\ (h^{-1})$ | 0.4372 ± 0.0002 | 0.437 ± 0.015 |
| $\beta_1\ (h^{-1})$ | 0.2500 ± 0.0005 | 0.25 ± 0.10 |
| $\beta_2\ (h^{-1})$ | 0.0001 ± 0.0003 | 0.000 ± 0.085 |

**Supplementary Table 6.** Results of the model parameters obtained by separate non-linear regression of each dataset: Mean number of radiation-induced foci per track $\bar{n}_Q$, fraction of persistent radiation-induced foci $p_Q$, mean number of persistent radiation-induced foci per track $\bar{p}_Q$ (calculated from $\bar{n}_Q$ and $p_Q$), repair rates $\beta_1$ and $\beta_2$, and respective standard errors (SE) obtained from the fit of the non-linear model (equation 3) to the difference of the data of a radiation quality and the sham-irradiated cells. The values are from the regression performed using the MPfit procedure of GDL.

| Radiation Quality | LET (keV/μm) | Mean number of foci per track, $\bar{n}_Q$ | Proportion of persistent radiation-induced foci, $p_Q$ | Mean number of persistent foci per track, $\bar{p}_Q$ | Repair rate, $\beta_1$ ($h^{-1}$) | Repair rate, $\beta_2$ ($h^{-1}$) |
|---|---|---|---|---|---|---|
| protons | | | | | | |
| 3 MeV | 19 ± 2 | 0.37 ± 0.04 | 0.11 ± 0.06 | *0.04 ± 0.02* | 0.34 ± 0.11 | 0.00 ± 0.00 |
| α – particles | | | | | | |
| 20 MeV | 36 ± 1 | 0.89 ± 0.16 | 0.38 ± 0.15 | *0.34 ± 0.15* | 0.96 ± 0.56 | 0.09 ± 0.04 |
| 10 MeV | 85 ± 4 | 1.53 ± 0.45 | 0.43 ± 0.19 | *0.66 ± 0.34* | 1.11 ± 0.92 | 0.09 ± 0.03 |
| 8 MeV | 170 ± 40 | 1.70 ± 0.47 | 0.27 ± 0.27 | *0.47 ± 0.48* | 0.43 ± 0.44 | 0.05 ± 0.06 |

**Supplementary Table 7.** Results of the model parameters obtained by separate non-linear regression of each dataset: Mean number of radiation-induced foci per track $\bar{n}_Q$, fraction of persistent radiation-induced foci $p_Q$, mean number of persistent radiation-induced foci per track $\bar{p}_Q$ (calculated from $\bar{n}_Q$ and $p_Q$), repair rate $\beta_1$, and respective standard errors (SE) obtained from the fit of the non-linear model (equation 3) to the difference of the data of a radiation quality and the sham-irradiated cells. The values are from the regression performed using the MPfit procedure of GDL imposing $\beta_2 = 0$.

| Radiation Quality | LET (keV/μm) | Mean number of foci per track, $\bar{n}_Q$ | Proportion of persistent foci, $p_Q$ | Mean number of persistent foci | Repair rate, $\beta_1$ ($h^{-1}$) |
|---|---|---|---|---|---|
| protons | | | | | |
| 3 MeV | 19 ± 2 | 0.37 ± 0.04 | 0.11 ± 0.06 | 0.04 ± 0.02 | 0.34 ± 0.11 |
| α – particles | | | | | |
| 20 MeV | 36 ± 1 | 0.73 ± 0.10 | 0.07 ± 0.04 | 0.07 ± 0.03 | 0.26 ± 0.06 |
| 10 MeV | 85 ± 4 | 1.17 ± 0.21 | 0.06 ± 0.04 | 0.07 ± 0.05 | 0.21 ± 0.06 |
| 8 MeV | 170 ± 40 | 1.62 ± 0.36 | 0.09 ± 0.05 | 0.15 ± 0.09 | 0.26 ± 0.08 |

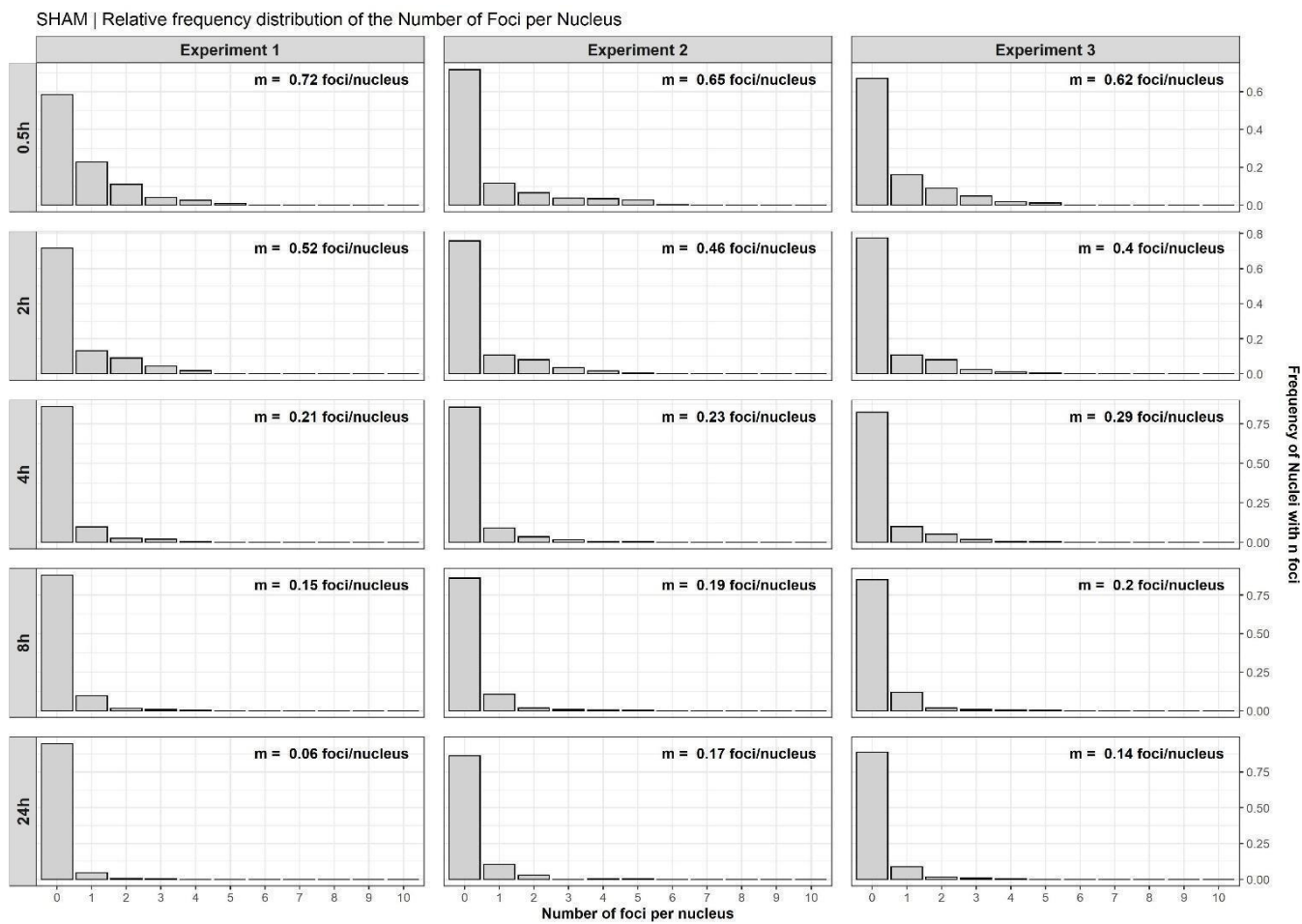

**Supplementary Figure 1.** Relative frequency distribution of the number of 53BP1 foci per nucleus for sham-treated cells, for three replicate experiments and for each timepoint after scanning. Inside each histogram, the mean number of foci per nucleus, *m*, is indicated.

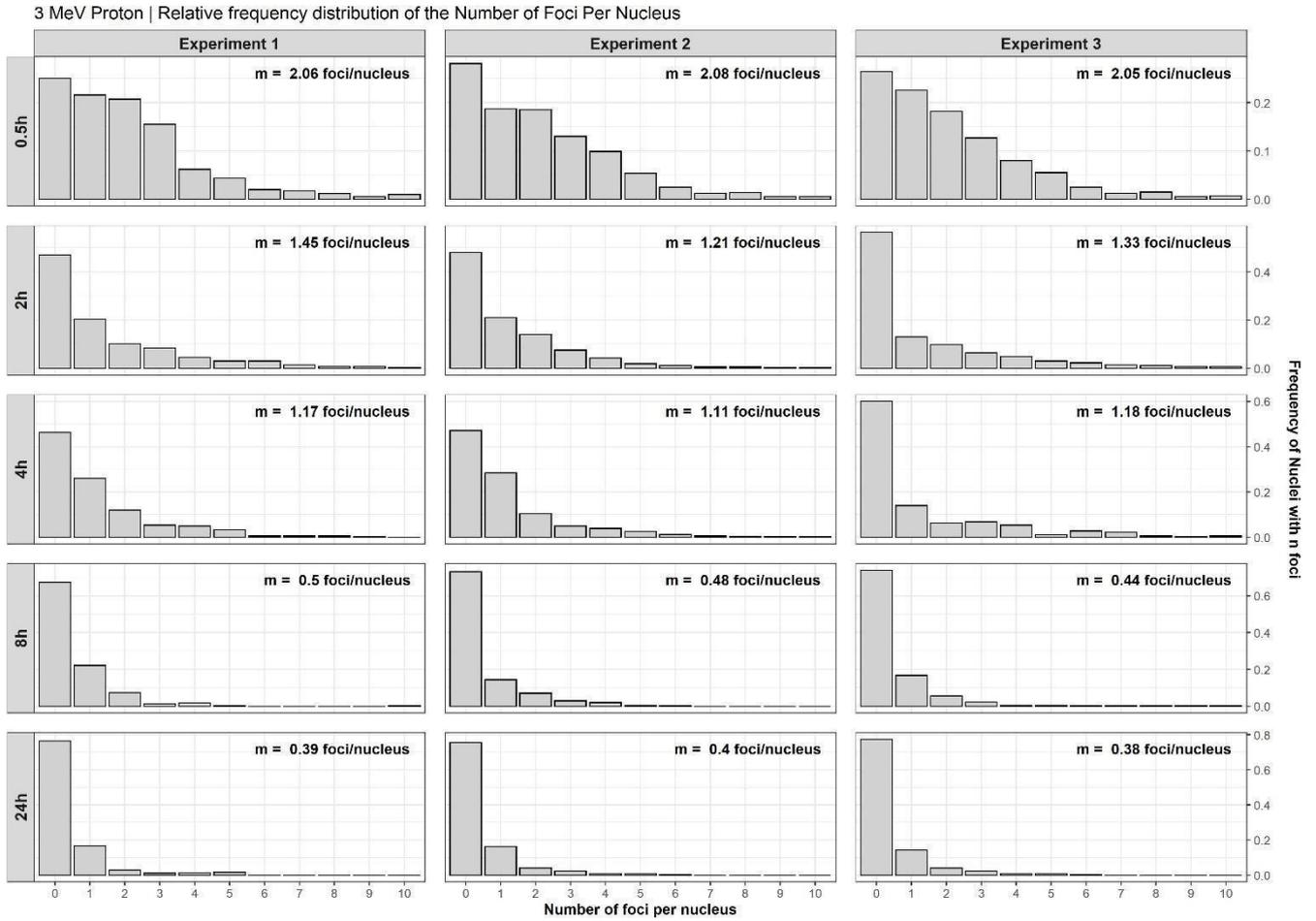

**Supplementary Figure 2.** Relative frequency distribution of the number of 53BP1 foci per nucleus for cells irradiated with 3 MeV Protons, for three replicate experiments and for each timepoint after irradiation. Inside each histogram, the mean number of foci per nucleus, *m*, is indicated.

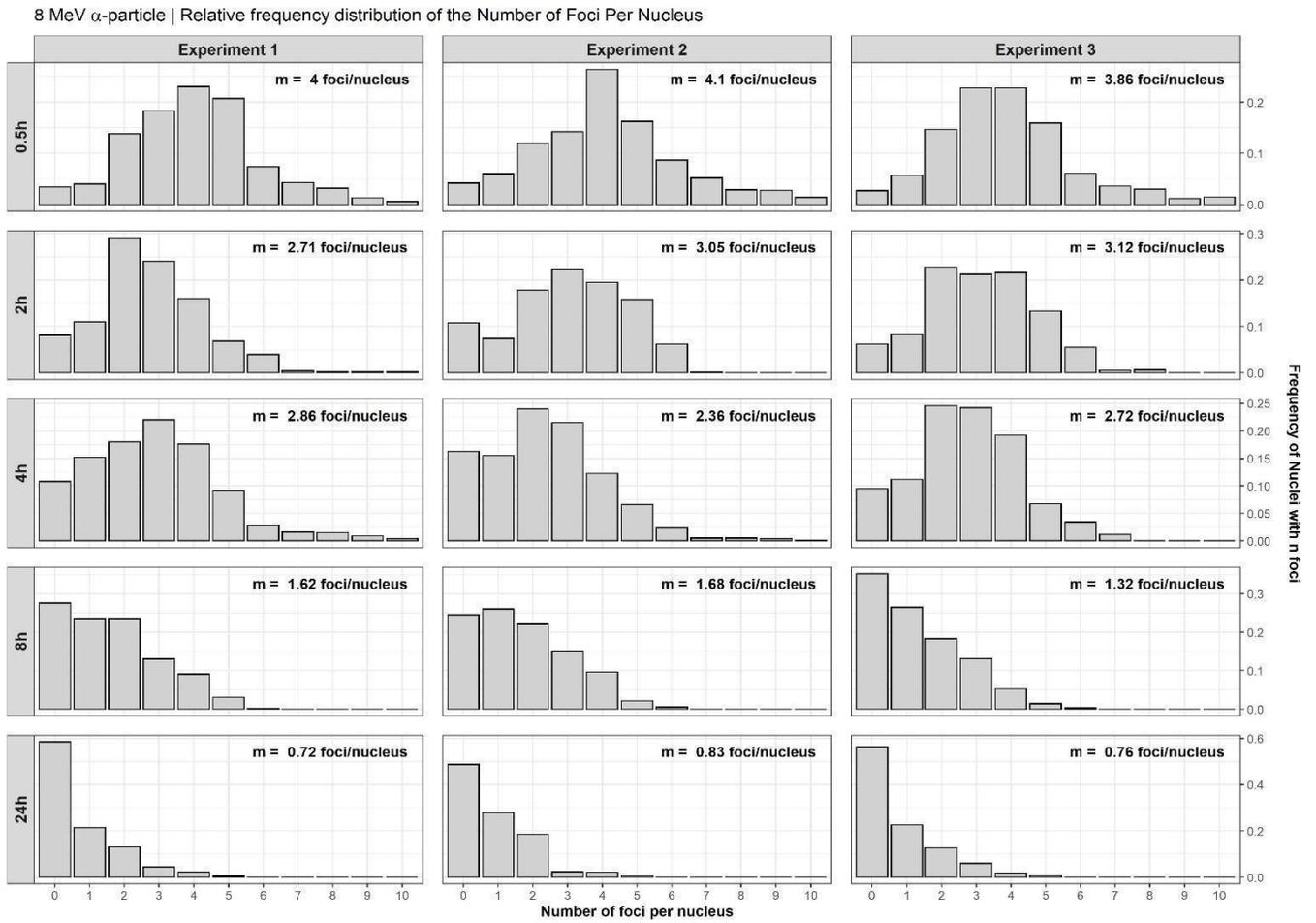

**Supplementary Figure 3.** Relative frequency distribution of the number of 53BP1 foci per nucleus for cells irradiated with 8 MeV α-particles, for three replicate experiments and for each timepoint after irradiation. Inside each histogram, the mean number of foci per nucleus, *m*, is indicated.

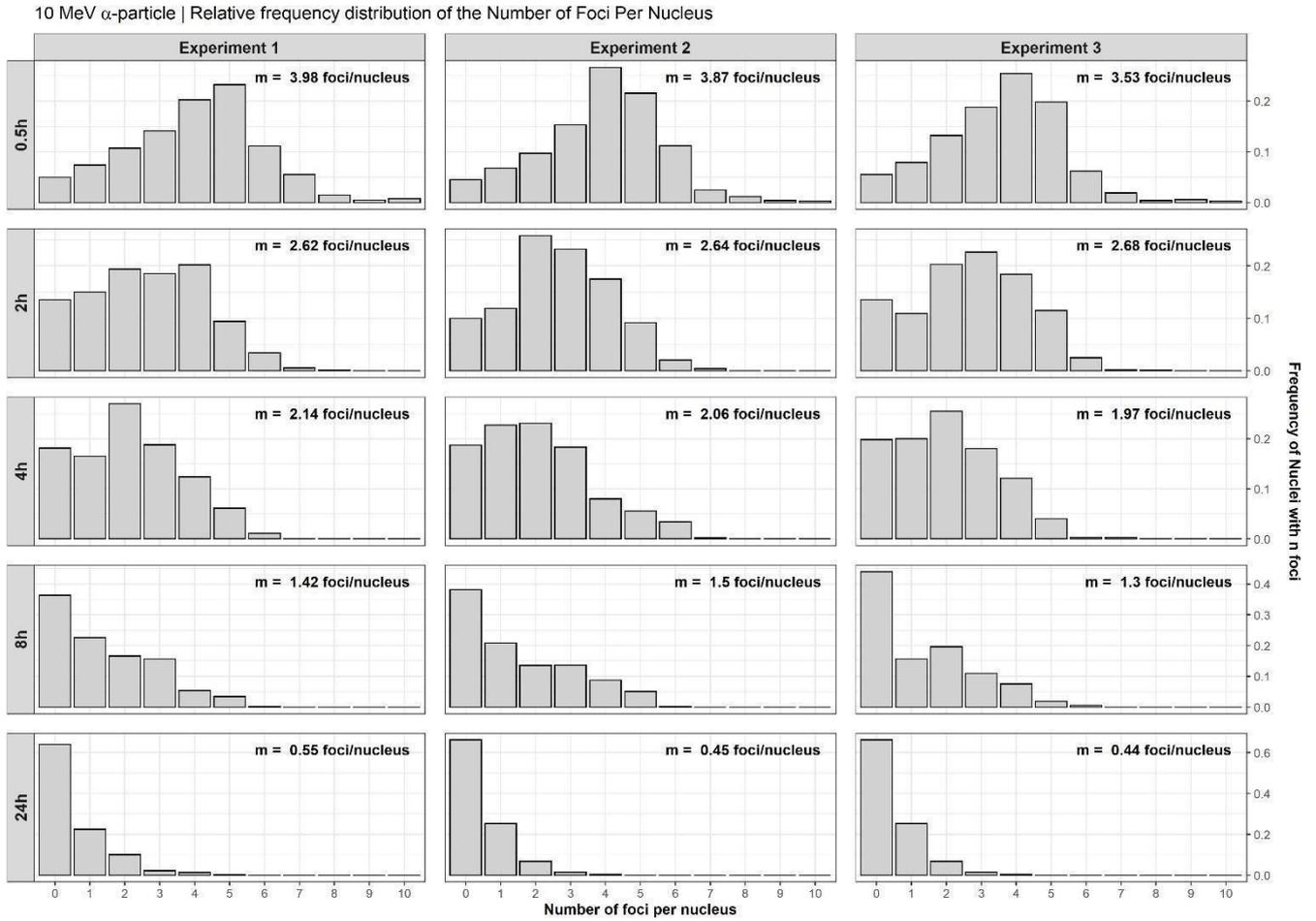

**Supplementary Figure 4.** Relative frequency distribution of the number of 53BP1 foci per nucleus for cells irradiated with 10 MeV α-particles, for three replicate experiments and for each timepoint after irradiation. Inside each histogram, the mean number of foci per nucleus, *m*, is indicated.

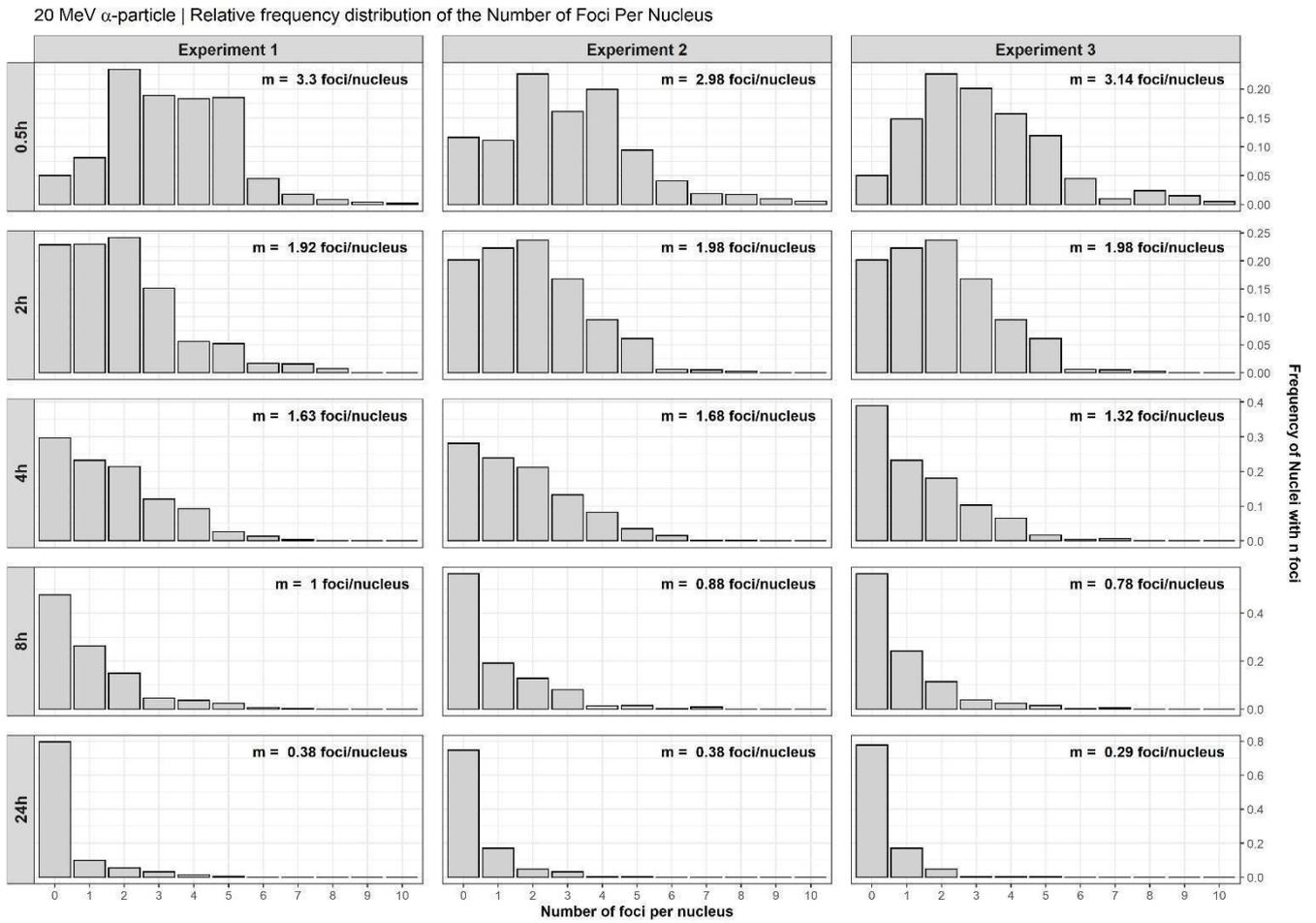

**Supplementary Figure 5.** Relative frequency distribution of the number of 53BP1 foci per nucleus for cells irradiated with 20 MeV α-particles, for three replicate experiments and for each timepoint after irradiation. Inside each histogram, the mean number of foci per nucleus, *m*, is indicated.